\documentclass[aps,prd,reprint,groupedaddress,amsmath,amssymb]{revtex4-2}

\usepackage{graphicx}
\usepackage{bm}
\usepackage{amssymb}
\usepackage{amsmath}
\usepackage{epsfig}
\usepackage{color}
\usepackage{mathtools}
\usepackage{cases}
\usepackage{booktabs}

\usepackage{dcolumn}
\usepackage{bm}
\usepackage{braket}

\usepackage[
colorlinks=true,
filecolor=black,
anchorcolor=blue,
linkcolor=blue,
citecolor=cyan, 
urlcolor=cyan,
linktocpage=true,
plainpages=false,
breaklinks=true,
pdfstartview=FitH
]{hyperref}

\newcommand{\mmn}{M_{\mu\nu}}
\newcommand{\mn}{\mu\nu}

\newcommand{\G}{{\cal{G}}}
\newcommand{\Psio}{\Psi_{\rm o}}
\newcommand{\calG}{{g}}
\newcommand{\mpl}{M_{\rm Pl}}
\newcommand{\tg}{\tilde{g}}

\begin{document}
\title{Angular correlation and deformed Hellings-Downs curve from\\ spin-2 ultralight dark matter}
\author{Rong-Gen Cai$^{2,3,1}$}
\email
{cairg@itp.ac.cn}
\author{Jing-Rui Zhang$^{1,3,4}$}
\email
{zhangjingrui22@mails.ucas.ac.cn}
\author{Yun-Long Zhang$^{5,1,6}$}
\email
{zhangyunlong@nao.cas.cn}
\affiliation{$^{1}$School of Fundamental Physics and Mathematical Sciences,  Hangzhou   Institute for Advanced Study, UCAS, Hangzhou 310024, China.}
\affiliation{$^{2}$School of Physical Science and Technology, Ningbo University, Ningbo, 315211, China}
\affiliation{$^{3}$CAS Key Laboratory of Theoretical Physics, Institute of Theoretical Physics, Chinese Academy of Sciences, Beijing 100190, China.} 
\affiliation{$^{4}$University of Chinese Academy of Sciences, Beijing 100190, China.}
\affiliation{$^{5}$National Astronomy Observatories, Chinese Academy of Science, Beijing, 100101, China}
\affiliation{$^{6}$International Center for Theoretical Physics Asia-Pacific, Beijing/Hangzhou, China
}

\begin{abstract}
The pulsar timings are sensitive to both the nanohertz gravitational-wave background and the oscillation of ultralight dark matter. The Hellings-Downs angular correlation curve provides a criterion to search for stochastic gravitational-wave backgrounds at nanohertz via pulsar timing arrays. We study the angular correlation of the timing residuals induced by the spin-2 ultralight dark matter, which is different from the usual Hellings-Downs correlation. At a typical frequency, we show that the spin-2 ultralight dark matter can give rise to the deformation of the Hellings-Downs correlation curve induced by the stochastic gravitational wave background.
\end{abstract}
\maketitle

\tableofcontents

\section{Introduction}
The detection of gravitational waves(GWs) has opened a new era for both theoretical and observational astronomy and cosmology~\cite{LIGOScientific:2016aoc}. In different frequency ranges, there exist several different approaches for the detection of GWs ~\cite{LIGOScientific:2014pky,VIRGO:2014yos,LISA:2017pwj,Ruan:2018tsw}. With the help of pulsar timing arrays(PTAs), we can detect the signals of stochastic gravitational wave background(SGWB) at frequencies around nanohertz. According to theoretical predictions, there are different physical processes that can contribute to SGWB, including supermassive black hole binaries~\cite{Burke-Spolaor:2018bvk}, domain walls~\cite{Zhang:2023nrs}, scalar curvature perturbations~\cite{Balaji:2023ehk}, etc. Compared with other physical effects which may also induce pulsar timing residuals, SGWB has the unique feature that the cross-correlation for different pulsars shows a Hellings-Downs pattern~\cite{Hellings:1983fr}, which can help the recognition of SGWB in data analysis.

Recently, several collaborations have reported their newest PTA data analysis independently~\cite{NANOGrav:2023gor,EPTA:2023fyk,Reardon:2023gzh,Xu:2023wog}, showing the evidence of detection for SGWB. In particular, the cross-correlation pattern resembles the Hellings-Downs correlation, which increases the reliability of the results.
On the other hand, it turns out that ultralight dark matter(ULDM) may also have effects on PTAs~\cite{Khmelnitsky:2013lxt,Nomura:2019cvc,Armaleo:2020yml,Sun:2021yra,Unal:2022ooa,Wu:2023dnp,Omiya:2023bio,Chowdhury:2023xvy}. ULDM has drawn much attention in recent years. Compared with cold dark matter, ULDM can suppress the structure formation on sub-galactic scales~\cite{Ferreira:2020fam}. The oscillation frequency of ULDM is their mass, and for ULDM with mass $m\sim10^{-22}\ \rm eV$, the corresponding frequency range is around nanohertz. Therefore, the oscillation of gravitational potential induced by ULDM can 
also induce pulsar timing residuals with a particular kind of angular dependence, which may contaminate the Hellings-Downs pattern. 

The spin nature of ULDM plays a vital role on its angular correlation on pulsar timing residuals. For example, the scalar ULDM has equal effects on pulsar at each direction~\cite{Khmelnitsky:2013lxt}, while the vector ULDM has strongly anisotropic behaviour~\cite{Nomura:2019cvc}. Here we focus on the spin-2 ULDM, which has gained much attention in recent years~\cite{Hassan:2011zd,Schmidt-May:2015vnx,Aoki:2016zgp,Babichev:2016hir,Babichev:2016bxi,Aoki:2017cnz,Marzola:2017lbt}. The theoretical origin of spin-2 ULDM comes from the bimetric theory~\cite{Hassan:2011zd}. In the bimetric theory, there are two kinds of gravitons in the mass spectrum, one of which is massless, and the other one is massive. This kind of massive graviton can be a dark matter candidate~\cite{Aoki:2016zgp,Babichev:2016hir,Babichev:2016bxi,Aoki:2017cnz,Marzola:2017lbt,Manita:2022tkl,Kolb:2023dzp,Gorji:2023cmz,Gialamas:2023aim,Gialamas:2023lxj,Guo:2023gfc}. 

The spin-2 ULDM has both gravitational effects~\cite{Wu:2023dnp} and coupling effects~\cite{Armaleo:2020yml,Sun:2021yra} on PTAs. In this paper, we focus on the coupling effects. We calculate the cross-correlation of pulsar timing residuals induced by spin-2 ULDM and the resulting angular pattern is purely quadrupole. We also show that certain parameter choices would cause the Hellings-Downs curve to be slightly or strongly deformed, so the PTA data may help constrain the parameter space.

This paper is organized as follows. In Section \ref{sec2}, we show the homogeneous background solution of the spin-2 ULDM. In Section \ref{sec3}, we calculate pulsar timing residuals induced by spin-2 ULDM. In Section \ref{sec4}, we evaluate the deformation of spin-2 ULDM on the Hellings-Downs curve, and we summarize the result in Section \ref{sec5}.

\section{Homogeneous background of spin-2 ULDM}\label{sec2}

 The spin-2 ULDM may come from the bimetric theory \cite{Hassan:2011zd, Schmidt-May:2015vnx,Babichev:2016hir,Aoki:2016zgp, Marzola:2017lbt}. We begin with the following action 
\begin{align}
S_{\rm spin-2}=\int d^4x \sqrt{|{\calG}|}\left[{\mpl^2}R({\calG})+{\cal L}_{M}\right],
\end{align}
where the Lagrangian density
\begin{align}
{\cal L}_{M}= -\frac{1}{2} M^{\mu\nu}{\cal E}^{~~\lambda\kappa}_{\mu\nu}M_{\lambda\kappa} 
 -\frac{m^2}{4}(M_{\mu\nu}M^{\mu\nu}-M^2),
\end{align}
and the Lichnerowicz operator is given by ${\cal E}^{~~\lambda\kappa}_{\mu\nu}\equiv-\frac{1}{2}( \delta^\lambda_\mu
\delta^\kappa_\nu \square-\calG_{\mu\nu}\calG^{\lambda\kappa}\square+\calG^{\lambda\kappa}\nabla_\mu\nabla_\nu +\calG_{\mu\nu}\nabla^\lambda\nabla^\kappa-2\nabla^\lambda\nabla_{(\mu}\delta^\kappa_{\nu)})$. 
Since the cosmic expansion is negligible on galactic scales and $m\gg H$, we assume the background metric to be flat, so that $\calG_{\mu\nu}=\eta_{\mu\nu}$.
The equation of motion for $\mmn$ from the Lagrangian density is 
\begin{align}
    {\cal E}^{~~\lambda\kappa}_{\mn} &M_{\lambda\kappa}
    +\frac{1}{2}m^2(\mmn-{\eta}_{\mn} M)=0 \label{eom}\ ,
\end{align}
where $M={\eta}^{\mu\nu}\mmn$. Applying $\partial^\mu$ on \eqref{eom} gives $\partial^\mu M_{\mu\nu}=\partial_\nu M$, and thus $\partial^\mu\partial^\nu M_{\mu\nu}=\square M$. Taking trace of \eqref{eom} and considering
$\eta^{\mu\nu}{\cal E}^{~~\lambda\kappa}_{\mn} M_{\lambda\kappa}=0$,  we can get $M=0$. Then the equation of motion  \eqref{eom} can be rewritten as $(\square-m^2)M_{\mu\nu}=0$. Together we obtain
\begin{equation}
    M=0,\ 
    \partial^\mu M_{\mu\nu}=0,\ (\square-m^2)M_{\mu\nu}=0 \label{eom1}\ .
\end{equation}
Since the typical velocity of dark matter in the galaxy is $v/c\sim 10^{-3}$, $M_{00}$ and $M_{0i}$ components are all suppressed and can be ignored.
The homogeneous background solution is then~\cite{Armaleo:2020yml}
\begin{align}
    M_{ij}=\frac{\sqrt{2\rho_{\mathrm{DM}}(\textbf{x})}}{m}\cos\left(mt+\varphi(\textbf{x}) \right) \varepsilon_{ij}(\textbf{x})\label{eq_mij}\ ,
\end{align}
where $\varepsilon_{ij}$ is the polarization tensor, $\rho_{\rm DM}({\bf x})$ is the energy density of dark matter at position ${\bf x}$, and $\varphi(\textbf{x})$ represents the phase. Since the occupation number in our galaxy is 
$\frac{\rho_{\rm DM}}{m\cdot(mv)^3}\approx 10^{95}\left(\frac{\rho_{\rm DM}}{0.4\rm GeV/cm^3}\right)\left(\frac{10^{-23}\rm eV}{m}\right)^4\left(\frac{10^{-3}}{v}\right)^3$,
the spin-2 field can be described by a classical wave, and note that the de Broglie wavelength for the spin-2 massive particle with mass $m$ is 
\begin{equation}
    \lambda_{\rm DM}=\frac{2\pi}{mv}\approx 4{\rm kpc} \left(
    \frac{10^{-23}{\rm eV}}{m}
    \right)
    \left(
    \frac{10^{-3}}{v}
    \right)\ ,
\end{equation}
so inhomogeneities within this distance can be smoothed out.

A way to parameterize $\varepsilon_{ij}$ is to separate it in terms of the spin states~\cite{Armaleo:2019gil,Armaleo:2020efr}
\begin{align}
    \varepsilon_{ij}=\sum_s \varepsilon_s \mathcal{Y}_{ij}^s\ ,
\end{align}
where $s$ represents the 5 different spin states $(\times,\ +,$ $\ \rm L,\ R,\ S)$, with
\begin{align}
    \varepsilon_\times& = \varepsilon_{\rm T}\cos\chi,\quad
    \varepsilon_+  = \varepsilon_{\rm T}\sin\chi,\\
    \varepsilon_{\rm L}& = \varepsilon_{\rm V}\cos\eta,\quad
    \varepsilon_{\rm R}  = \varepsilon_{\rm V}\sin\eta\ .
\end{align}
Here $\chi$ and $\eta$ determine the azimuthal direction of the tensor part and vector part of the spin-2 field, respectively. Moreover, $\epsilon_{\mu\nu}$ is normalized such that
\begin{align}
    \sum_s\varepsilon_s^2 =1\ .
\end{align}
The corresponding tensors are
\begin{align}
    \mathcal{Y}^\times_{ij}&=\frac{1}{\sqrt{2}}(p_iq_j+q_ip_j)\ ,\\
    \mathcal{Y}^+_{ij}&=\frac{1}{\sqrt{2}}(p_ip_j-q_iq_j)\ ,\\
    \mathcal{Y}^{\rm L}_{ij}&=\frac{1}{\sqrt{2}}(q_ir_j+r_iq_j)\ ,\\
    \mathcal{Y}^{\rm R}_{ij}&=\frac{1}{\sqrt{2}}(p_ir_j+r_ip_j)\ ,\\
    \mathcal{Y}^{\rm S}_{ij}&=\frac{1}{\sqrt{6}}(3r_ir_j-\delta_{ij})\ ,
\end{align}
which represent {-2, 2, -1, 1, 0} spin states, respectively, $r_i$ is the unit vector with propagation direction of the massive spin-2 field, and $p_i,q_i$ are two unit vectors orthogonal to $r_i$ and to each other. Note that in general, spin-2 fields can have at most 6 degrees of freedom, while in the case of bimetric theory~\cite{Marzola:2017lbt}, the traceless condition restricts the number to 5.

\section{Pulsar timing residuals induced by spin-2 ULDM}\label{sec3}

To study the effects on the PTA signal, a convenient way to calculate is to change the frame as in~\cite{Armaleo:2020yml},
$\tg_{\mu\nu}={\calG}_{\mu\nu}-\frac{\alpha}{M_{\rm Pl}}M_{\mu\nu}$, where $\alpha$ is the coupling constant characterizing the strength of the direct coupling between normal matter and the spin-2 ULDM.
Then we evaluate the geodesics of photons based on the new metric $\tg_{\mu\nu}$. Here we assume a flat background since $m\gg H$ and work with $\tg_{ij}=\delta_{ij}-\frac{\alpha}{M_{\rm Pl}}M_{ij}$.
For a photon travelling along the geodesic from the pulsar $a$ to the Earth with 4-momentum $p^\mu=\nu(1,n_a^i)$, the geodesic equation gives
\begin{align}
     \frac{dp^0}{{du}}=\frac{\alpha\nu^2}{2M_{\rm Pl}}\partial_t M_{ij}n_a^in_a^j\ ,
\end{align}
where $u$ is the affine parameter. Treating $\alpha$ as a small parameter and keeping first order terms in $\alpha$, we get 
\begin{align}
     \nu={\bar\nu}\left(
    1+\frac{\alpha}{2M_{\rm Pl}}\int^{\rm E}_a
    {du}{\bar\nu}\partial_t M_{ij}n_a^in_a^j
    \right)\ ,
\end{align}
where the integral is evaluated from the pulsar $a$ to the Earth(E), and ${\bar\nu}$ is the frequency of the photon if not perturbed. 

By making use of ${\bar\nu}\partial_t=\tfrac{d}{{du}}-{\bar\nu} n_a^i\partial_i$, it becomes
\begin{align}
      \nu ={\bar\nu}\Big(&
    1+\frac{\alpha}{2M_{{\rm Pl}}}(M_{ij}^{\rm E}-M_{ij}^{a})n_a^in_a^j-\nonumber\\
    &\frac{\alpha}{2M_{{\rm Pl}}}
    \int^{{\rm E}}_{a}
    {du}\ \nu_0n^k \partial_k M_{ij}n_a^in_a^j
    \Big)\ .
\end{align}
Compared with the second term, the last term can be ignored since it contains space derivatives which introduce a factor $v/c\sim {\cal O}(10^{-3})$. The resulting redshift $z_a=({\bar\nu}_{a}-\nu_a)/{\bar\nu}_{a}$ for signals from the pulsar a is
\begin{align}
   z_a&=-\frac{\alpha}{2M_{{\rm Pl}}}(M_{ij}^{\rm E}-M_{ij}^{a})n_a^in_a^j\nonumber\\
   &=F_{a}^{\rm E}\Psio({\bf x}_{\rm E})\cos(mt+\varphi_{\rm E})\nonumber\\
   &\quad-F_{a}^{\rm P}\Psio({\bf x}_{\rm a})\cos(mt-mL_{a}+\varphi_{a})\ ,
\end{align}
where $L_{a}$ is the distance between the pulsar $a$ and the Earth, and the corresponding beam pattern functions are
\begin{align}
    F_{a}^{\rm E}&=-\varepsilon_{ij}({\bf x}_{\rm E})n_a^in_a^j\ ,\\
    F_{a}^{\rm P}&=-\varepsilon_{ij}({\bf x}_{\rm a})n_a^in_a^j\ ,
\end{align}
which contain the dependence of $z_{\rm a}$ on angles, while 
\begin{align}
    \Psio({\bf x}_{\rm E})&=\frac{\alpha\sqrt{\rho_{\rm DM}({\bf x}_{\rm E})}}{\sqrt{2}m M_{\rm Pl}}\label{eq_psi},\\
    \Psio({\bf x}_{\rm a})&=\frac{\alpha\sqrt{\rho_{\rm DM}({\bf x}_{\rm a})}}{\sqrt{2}m M_{\rm Pl}},
\end{align}
represent the amplitude of $z_{\rm a}$, which depends on local energy density of dark matter.

The timing residuals
\begin{align}
    \Delta T_a(t)=\int^t_0 dt' z_a(t')
\end{align}
are then integrated to be
\begin{align}
    \Delta T_a(t)= &\frac{1}{m}\big[F_{a}^{\rm E}\Psio({\bf x}_{\rm E})\sin (mt+\varphi_{\rm E})\nonumber\\
    &-F_{a}^{\rm P}\Psio({\bf x}_{\rm a})\sin (mt-mL_a+\varphi_{a})\big]\ ,\label{tdm2}
\end{align}
with the origin of time chosen to cancel constant terms for simplicity.

\section{Angular correlation curve}\label{sec4}
For pulsar $a$ and pulsar $b$, the cross-correlation between their timing residuals can be written as
\begin{align}
    C_{ab}(\tau)=\langle\Delta T_a(t)\Delta T_b (t+\tau)\rangle-\langle\Delta T_a(t)\rangle \langle \Delta T_b (t+\tau)\rangle\ .
\end{align}
For the stochastic gravitational wave background, the cross-correlation it induces can be described by~\cite{Hellings:1983fr,Omiya:2023bio}
\begin{align}
    C^{\rm GW}_{ab}(\tau)=\sum_i \Gamma_{\rm HD}(\zeta)\Phi_{\rm GW}(f_i)\cos 2\pi f_i \tau\ ,
\end{align}
where $\Gamma_{\rm HD}$ describes the Hellings-Downs curve
\begin{align}
    \Gamma_{\rm HD}=&\frac{1}{2}-
    \frac{1}{4}\Big(\frac{1-\cos\zeta}{2}\Big)+\nonumber\\
    &+\frac{3}{2}\Big( \frac{1-\cos\zeta}{2}\Big)\ln \Big(\frac{1-\cos\zeta}{2}\Big)\ ,
\end{align}
and
\begin{align}
    \Phi_{\rm GW}(f_i)=\frac{1}{12\pi^2f_i^3}\frac{1}{T_{\rm obs}}h^2_c(f_i)\ .
\end{align}
The frequency integral is discretized based on the observation time $T_{\rm obs}$~\cite{NANOGrav:2023gor}, and $h_c$ is the characteristic
strain of SGWB, which usually can be parameterized as 
\begin{align}
    h_c(f)=A_{\rm GW}\left(\frac{f}{f_0}
    \right)^{\alpha_{\rm GW}}\ .
\end{align}

\subsection{Angular correlation of spin-2 ULDM}

For the spin-2 ULDM, the angular correlation 
$\langle\Delta T_a(t)\Delta T_b (t+\tau)\rangle=\frac{1}{T_{\rm obs}}\int_0^{T_{\rm obs}} \Delta T_a(t)\Delta T_b (t+\tau) dt $
can be obtained by using the timing residuals in \eqref{tdm2}. Notice that, since the observational period for PTAs is $T_{\rm obs}\gtrsim 10 {\rm yr}$, and the oscillation period for ultralight dark matter with mass $m\gtrsim 10^{-24}{\rm eV}$ is $m^{-1}\lesssim T_{\rm obs}$.
Thus, when we take the time average over  $T_{\rm obs}$, oscillating terms with $\cos(mt)$ can be ignored. The dominant cross-correlation terms are
\begin{align}
    &\langle\Delta T_a(t)\Delta T_b (t+\tau)\rangle\nonumber\\
=& \frac{1}{2m^2}
   \langle F_{a}^{\rm E}F_{b}^{\rm E}\rangle
   \langle \Psio^2({\bf x}_{\rm E})\rangle
    \cos(m\tau)   \nonumber \\
+&\frac{1}{2m^2}
    \langle F_{a}^{\rm P}F_{b}^{\rm P}\rangle
    \langle\Psio({\bf x}_{a})
    \Psio({\bf x}_{b})\rangle
   \nonumber \\
&\quad \times\cos(m\tau+mL_a-mL_b-\Upsilon_a+\Upsilon_b) \nonumber\\
-&\frac{1}{2m^2}
    \langle F_{a}^{\rm E}F_{b}^{\rm P}\rangle
    \langle\Psio({\bf x}_{\rm E})
    \Psio({\bf x}_{b})\rangle\nonumber\\
&\quad \times\cos(m\tau-mL_b+\Upsilon_b-\Upsilon_{\rm E}) \nonumber\\
-&\frac{1}{2m^2}
    \langle F_{a}^{\rm P}F_{b}^{\rm E}\rangle
    \langle\Psio({\bf x}_{\rm E})
    \Psio({\bf x}_{a})\rangle\nonumber\\
&\quad \times\cos(m\tau+mL_a+\Upsilon_{\rm E}-\Upsilon_{a})\ .
\end{align}
On the other hand, most of the terms are associated with the distance between the pulsars and the earth, which ranges from 0.1 kpc to 10 kpc. Then, for typical ULDM with $m\simeq 10^{-24}\rm eV$, these terms have a contribution to the phase as $10\leq m L_a\simeq m L_b\leq 1000$, which is far more than $2\pi$ and can be treated as random noise and averaged out.
Thus, the final result is
\begin{align}
    C^{\rm spin-2}_{ab}(\tau)\approx \frac{1}{2m^2}
   \langle F_{a}^{\rm E}F_{b}^{\rm E}\rangle
   \langle \Psio^2({\bf x}_{\rm E})\rangle
    \cos(m\tau)\ .
\end{align}

\begin{figure}[h]
\includegraphics[keepaspectratio, scale=0.36]{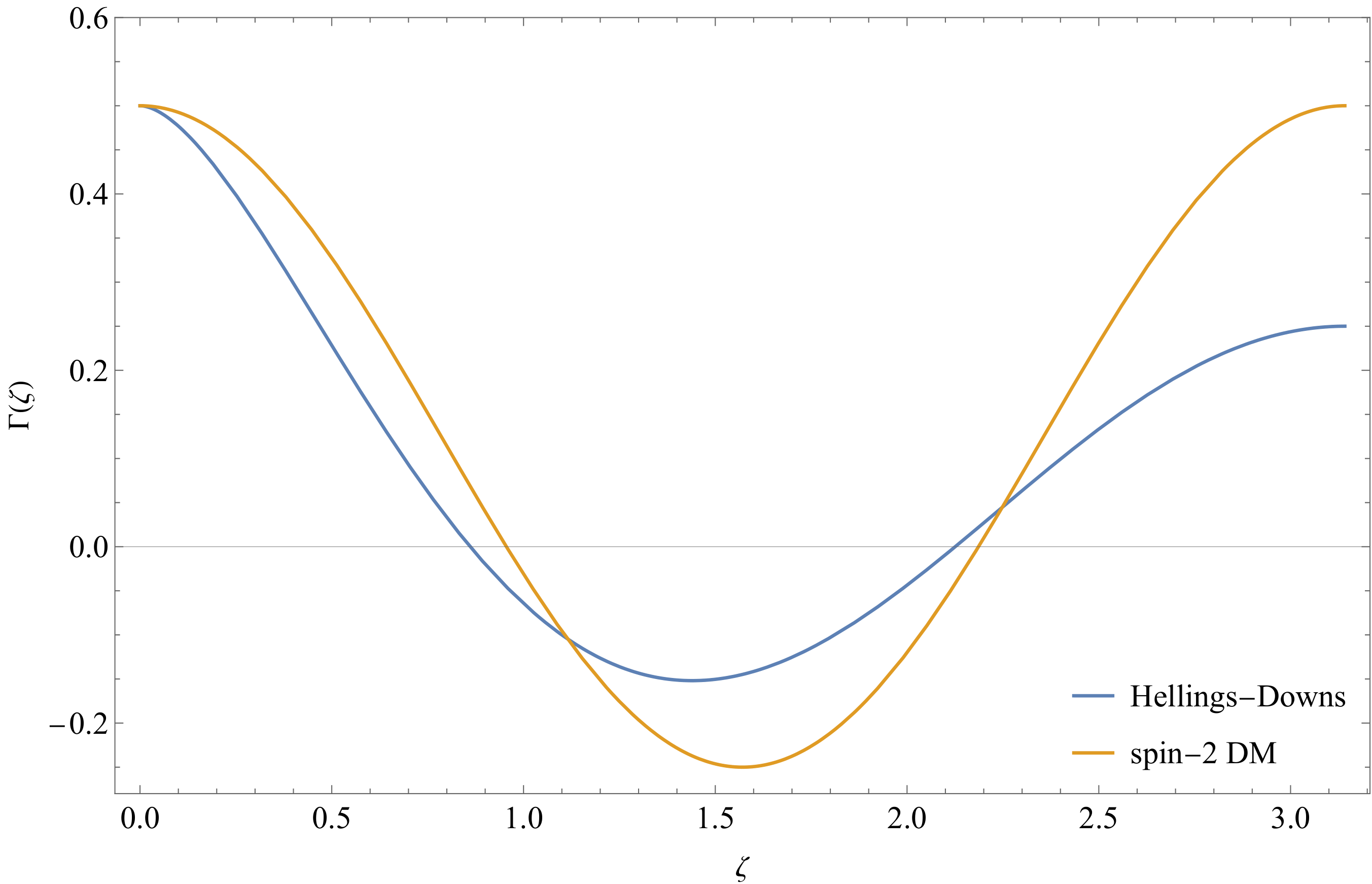}
\caption{The blue and orange curves represent the angular correlation of the timing residuals induced by stochastic GW background (Hellings-Downs curve) and spin-2 ULDM, respectively. Both curves are normalized at $\zeta=0$ such that $\Gamma(0)=1/2$.} 
\label{fig:HD1}
\end{figure}

The angular dependence of $C^{\rm spin-2}_{ab}$ is contained in $\langle F_{a}^{\rm E}F_{b}^{\rm E}\rangle$, we set the coordinate frame in which $n_a^i=(0,0,1)$ and $n_b^i=(\sin\zeta,0,\cos\zeta)$. Since observable pulsars are distributed all over the sky, effectively, we can assume the polarization for the massive spin-2 field as isotropic and integrate over possible directions~\cite{Omiya:2023bio} and also integrate over azimuthal directions of tensor and vector parts. For convenience, we set
\begin{align}
     \vec{r}&=(\sin\theta\cos\phi,\sin\theta\sin\phi,\cos\theta)\ ,\\
    \vec{p}&=(\sin\phi,-\cos\phi,0)\ ,\\
    \vec{q}&=(\cos\theta\cos\phi,\cos\theta\sin\phi, -\sin\theta)\ .
\end{align}
The Hellings-Downs-like curve for massive spin-2 background can be derived from $\varepsilon_{ij}n^in^j$,
\begin{align}
    \langle &F_{a}^{\rm E}F_{b}^{\rm E}\rangle\nonumber=
    \int\frac{d\eta}{2\pi} \frac{d\chi}{2\pi} \frac{d^2\Omega}{4\pi}
    \sum_s \varepsilon_s \mathcal{Y}_{ij}^s n_a^in_a^j
    \sum_\lambda \varepsilon_\lambda \mathcal{Y}_{kl}^\lambda n_b^kn_b^l\nonumber\\
    &= \frac{1}{30}(1+3\cos2\zeta)=\frac{4}{15}\Gamma_{\rm DM}(\zeta)\ ,
\end{align}
where
\begin{align}
    \Gamma_{\rm DM}(\zeta)
=\frac{1}{2}P_2(\cos\zeta)=\frac{1}{8}(1+3\cos2\zeta)
\end{align}
is the angular correlation curve normalized to be compared with Hellings-Downs curve in Figure \ref{fig:HD1}. Here we can see that the effect of spin-2 ULDM follows a purely quadrupole pattern.

The total result for the correlation is
\begin{align}
    C^{\rm DM}_{ab}(\tau)=\Phi_{\rm DM}\Gamma_{\rm DM}(\zeta)\cos(m\tau)\ ,
\end{align}
where $\Phi_{\rm DM}$ denotes the amplitude of the correlation
\begin{align}
    \Phi_{\rm DM}=\frac{2}{15m^2}\langle \Psio^2({\bf x}_{\rm E})\rangle\ .\label{eq_phi}
\end{align}

\subsection{Deformation of Hellings-Downs curve}

When SGWB and spin-2 ULDM are both considered, the total cross-correlation is 
\begin{align}
    C_{ab}(\tau)=&\sum_i \Gamma_{\rm HD}(\zeta)\Phi_{\rm GW}(f_i)\cos 2\pi f_i \tau\nonumber\\
    &\quad +\Gamma_{\rm DM}(\zeta)\Phi_{\rm DM} \cos(2\pi f_{m}\tau)\ ,
\end{align}
where the effect of spin-2 ULDM at frequency $f_{m}=m/2\pi$ can be incorporated into $\Gamma(\zeta)$ as~\cite{Omiya:2023bio}
\begin{align}\label{Gammaeff}
    \Gamma_{\rm eff}(\zeta)=&\frac{\Phi_{\rm GW}(m/2\pi)}{\Phi_{\rm GW}(m/2\pi)+\Phi_{\rm DM}}\Gamma_{\rm HD}(\zeta)+\nonumber\\
    &\frac{\Phi_{\rm DM}}{\Phi_{\rm GW}(m/2\pi)+\Phi_{\rm DM}}\Gamma_{\rm DM}(\zeta)\ .
\end{align}
In the case of SGWB, according to the observation of NANOGrav reported in~\cite{NANOGrav:2023gor}, for $\gamma=-2\alpha_{\rm GW} +3=13/3$, $A_{\rm GW}=2.4\times10^{-15}$ with the observation time $T_{\rm obs}\sim 15\  \rm yr$ and $f_{\rm ref}=1\  \rm yr^{-1}$, so the amplitude of the correlation is
\begin{align}
    &\Phi_{\rm GW}({m/2\pi})\sim 1\times 10^{-32}{\rm yr}^2\left(
    \frac{m}{10^{-22}{\rm eV}}\right)^{-\frac{13}{3}}\left(\frac{15{\rm yr}}{T_{\rm obs}}\right).
\end{align}

\begin{figure}[h]
\includegraphics[keepaspectratio, scale=0.36]{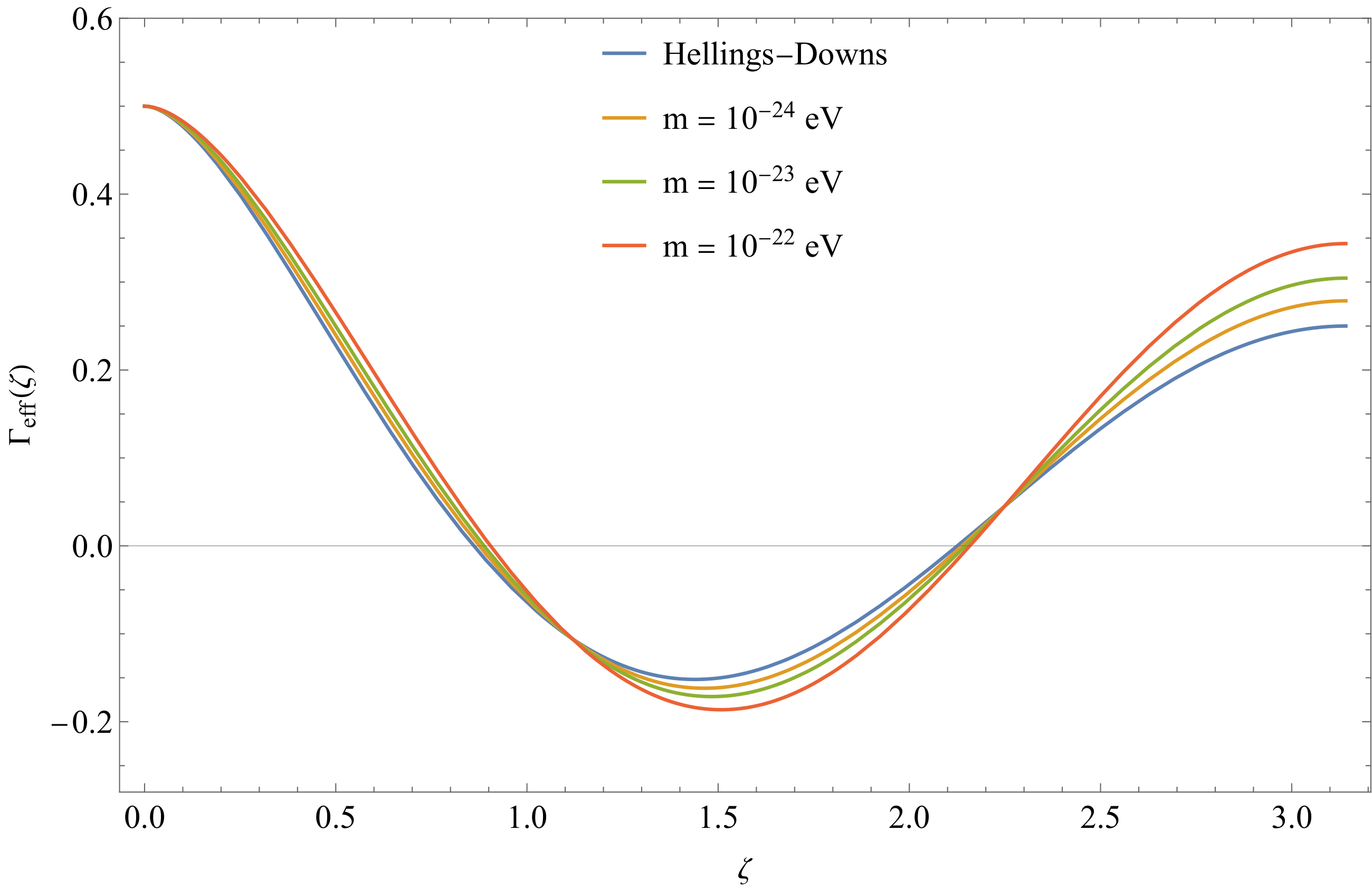}
\caption{Effective cross-correlation curves with $\alpha=10^{-6}$ and mass ranging from $10^{-24}$eV to $10^{-22}$eV. It can be seen that in this range and at the typical frequencies $f_{m}=m/2\pi$, the deformation of spin-2 ULDM on the Hellings-Downs curve is relatively small.} 
\label{fig:eff10-6}
\end{figure}

For the effects of ultralight dark matter, the amplitude of correlation can be derived from 
\eqref{eq_psi} and \eqref{eq_phi} as
\begin{align}
    \Phi_{\rm DM}=\frac{\alpha^2}{15m^4 }
    \frac{\rho_{\rm DM}({\bf x}_{\rm E})}{{\mpl^2}}\ ,
\end{align}
so that
\begin{align}
    \Phi_{\rm DM}\sim & \ 6 \times 10^{-33}{\rm yr}^2 \left(
    \frac{\rho_{\rm DM}}{0.4 \rm GeV/ cm^{3}}
    \right)\nonumber\\
    &\times\left(
    \frac{\alpha}{10^{-6}}
    \right)^2
    \left(
    \frac{m}{10^{-22}{\rm eV}}\right)^{-4}\ .
\end{align}

\begin{figure}
\includegraphics[keepaspectratio, scale=0.36]{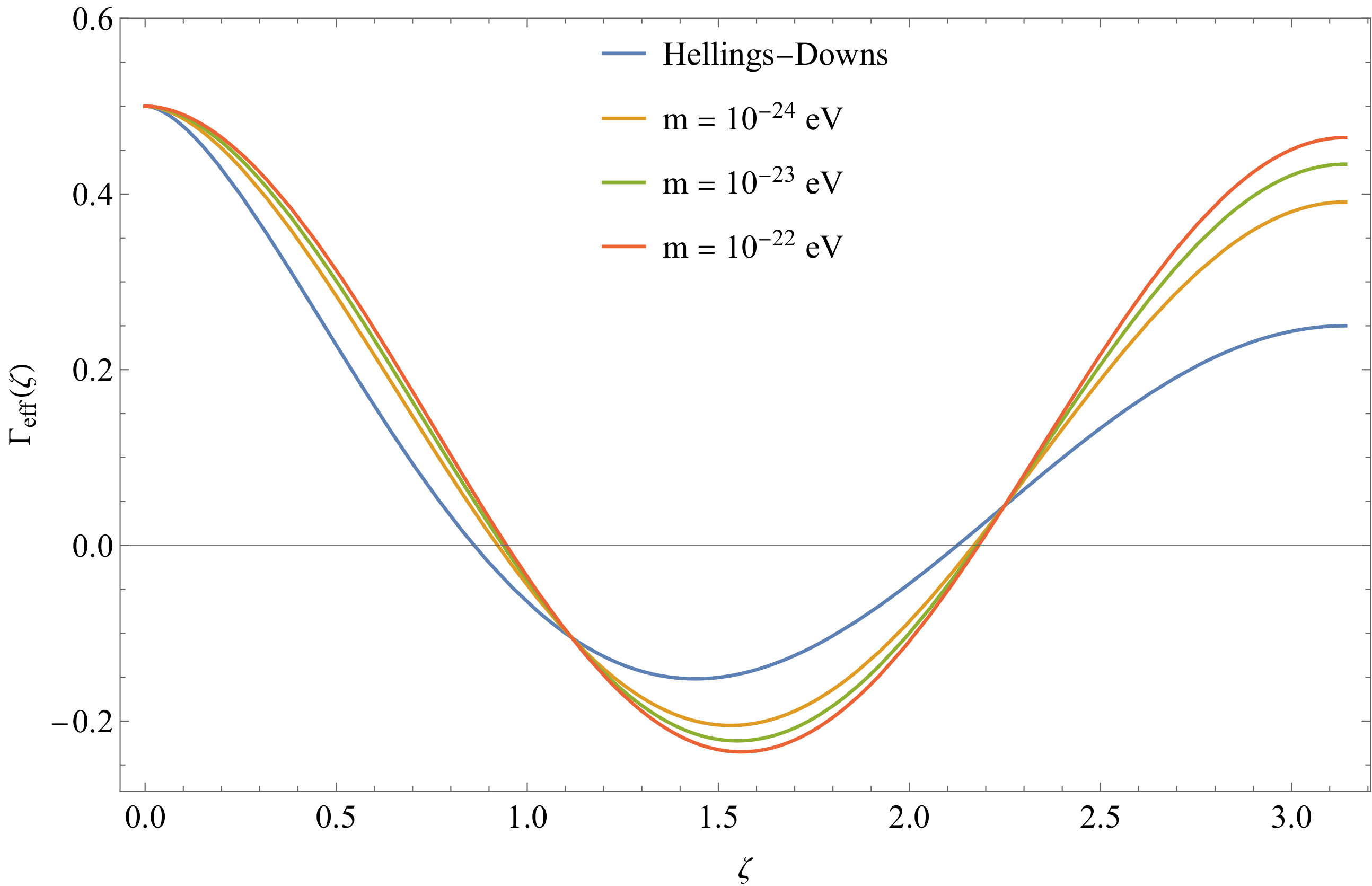}
\caption{Effective cross-correlation curves with $\alpha=10^{-5.5}$ and mass ranging from $10^{-24}$eV to $10^{-22}$eV. The deformation is very strong in this parameter range, suggesting that if the coupling constant $\alpha$ is above this magnitude, existing ultralight spin-2 ULDM would have considerable effects on the deformation of the Hellings-Downs curve at the typical frequencies $f_{m}=m/2\pi$.} 
\label{fig:eff10-5.5}
\end{figure}

\begin{figure}
\includegraphics[keepaspectratio, scale=0.36]{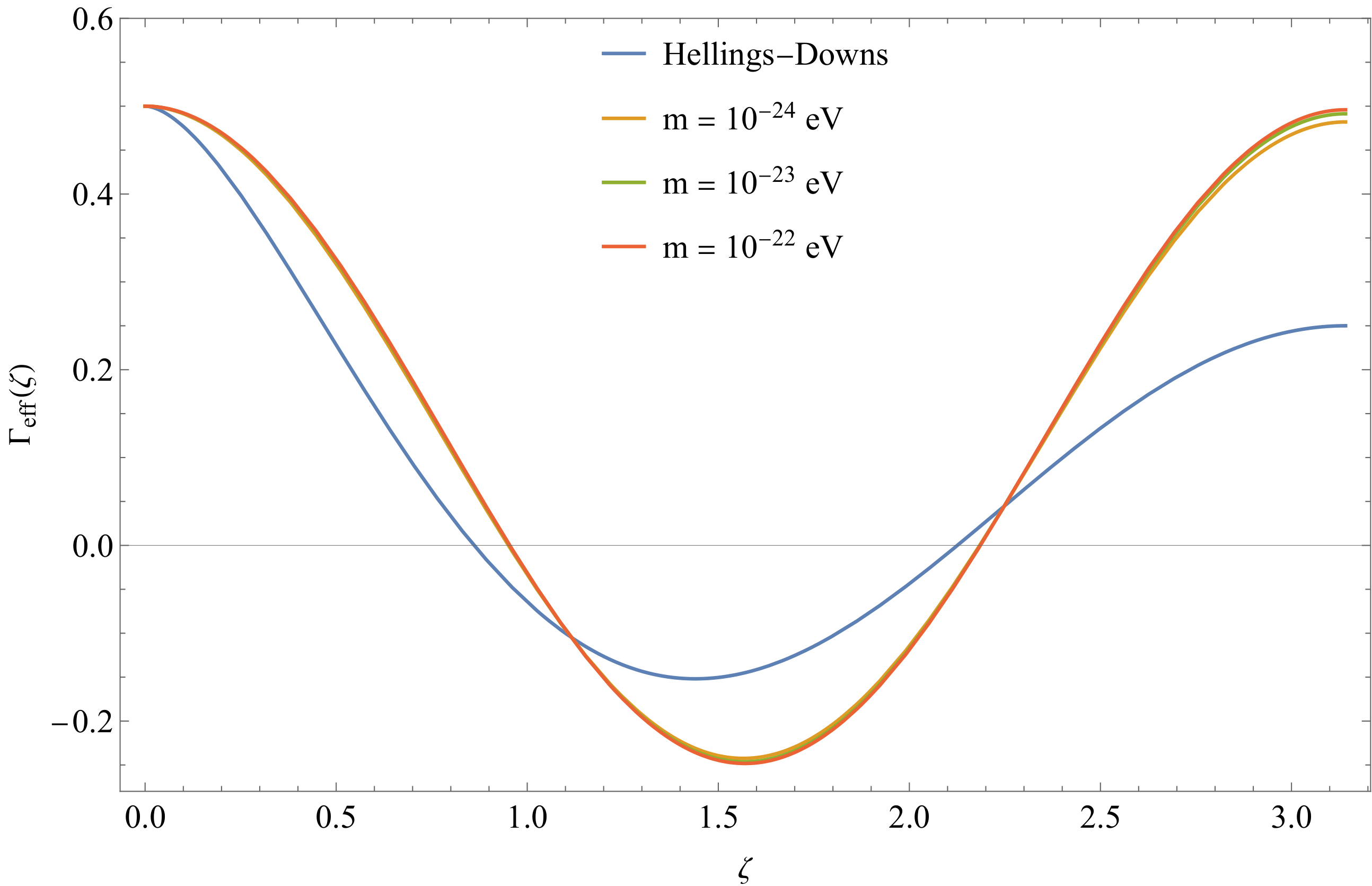}
\caption{Effective cross-correlation curves with $\alpha=10^{-5}$ and mass ranging from $10^{-24}$eV to $10^{-22}$eV. In this range of $\alpha$, the curves are dominated by the spin-2 ULDM at the typical frequencies $f_{m}=m/2\pi$.} 
\label{fig:eff10-5}
\end{figure}

Different choices of the parameters $\alpha$ and $m$ give different deformations of the Hellings-Downs curve, shown in Fig.~\ref{fig:eff10-6}, Fig.~\ref{fig:eff10-5.5} and Fig.~\ref{fig:eff10-5}. It can be seen from these figures that, compared with the mass of dark matter $m$, the order of magnitude of the coupling constant $\alpha$ has a significant impact on the deformation of the Hellings-Downs curve at certain frequencies. For example, for $\alpha>10^{-5}$, the shape of the curve is almost determined by the spin-2 ULDM. It is also interesting that with larger mass parameter of the ultralight dark matter, the deformation is larger and easier to be observed.

\section{Conclusion and Discussion}\label{sec5}
In summary, we show that spin-2 ULDM can induce the deformation of Hellings-Downs curve at the frequency of the mass parameter. Since the residuals induced by spin-2 ULDM is monochromatic, the deformation of Hellings-Downs curve is expected to happen only in a narrow frequency range around $f_m=m/2\pi$. Both the mass of spin-2 ULDM and the coupling constant $\alpha$ have impact on the deformation, so the observational data of PTAs can help to constrain the parameter space.

Compared with the spin-1 case in~\cite{Omiya:2023bio}, the spin-2 ULDM has a pure quadrupole effect on the correlation, which is expected from its tensorial nature. According to PTA data~\cite{Wu:2023dnp}, the coupling constant $\alpha$ is constrained at $10^{-6}\sim10^{-5}$ for mass $m<5\times 10^{-23}\rm eV$. The constraint is set merely according to the amplitude of gravitational strain. The angular correlation we have found gives a more concrete understanding on how the spin-2 ULDM can affect the PTA signal, causing angular correlation of the timing residuals, and deforming the correlation curve of stochastic gravitational wave background.

The spin-2 ULDM has different effects on pulsar timing residuals, compared with the stochastic gravitational wave background. 
The PTA residuals are induced by the local homogeneous background from pulsars to the Earth, and the Hellings-Downs curve is caused by isotropic gravitational waves coming from all over space.
The spin-2 ULDM is massive, which makes the frequency of the homogeneous background fixed to its mass. 
For  gravitational wave background from gravity theories beyond general relativity, the Hellings-Downs curves are also deformed but in a broad frequency region
\cite{Qin:2020hfy,Liang:2021bct,Bernardo:2022rif,Bernardo:2023mxc,Liang:2023ary}.
However, the deformation in our case is different from them.
In \cite{Qin:2020hfy,Liang:2021bct,Bernardo:2022rif,Bernardo:2023mxc,Liang:2023ary}, the deformation is present in all frequency ranges. In our case, the deformation is only in a narrow frequency range, which corresponds to the mass of the spin-2 ULDM. In addition, our angular correlation function is different from them. In their case, extra polarization modes are considered for the gravitational waves. Therefore, in the calculation of integrating over all directions, terms like $1/[(1+\hat{n}\cdot \hat{n}_a)(1+\hat{n}\cdot \hat{n}_b)]$ and $\left(1-e^{-2\pi if \tau_a(1+\hat{n}\cdot\hat{n}_a)}\right)\left(1-e^{2\pi if\tau_b(1+\hat{n}\cdot\hat{n}_b)}\right)$ need to be carefully treated for longitudinal or mixed polarization modes. 
In our case, such terms don't exist, because the spin-2 ULDM acts as a background to influence pulse signals coming from pulsars.

The deformed Hellings-Downs curve can be used to give an intuitive comparison with observational data and estimate the constraint on spin-2 ULDM. If apparent deformation of Hellings-Downs curve is observed in a certain frequency range, it can be compared with our results to verify whether the deformation is caused by the spin-2 ULDM. In this sense, PTAs act as a distinct dark matter detector for spin-2 ULDM. 
In~\cite{Nomura:2019cvc},  the angular correlation of spin-2 ULDM for different modes are obtained with several undetermined coefficients. In our case, we use the method of derivation for the Hellings-Downs curve, and all the coefficients are summed up. Thus, our result can be directly compared with the PTA data.
In~\cite{Wu:2023dnp}, the pure gravitational effects of the spin-2 ULDM are considered, which is different from our case on the coupling effects of $\alpha$ in spin-2 ULDM. In other words, the PTA effect they have discussed is the gravitational oscillation caused by the energy-momentum tensor of spin-2 ULDM. In our case, the direct coupling between matter and spin-2 ULDM is considered.
It is also interesting to analyze other effects of spin-2 ULDM e.g. in the solar system astrometry that was proposed more recently \cite{Mentasti:2023gmr}.

From the references~\cite{Bernardo:2022rif,Bernardo:2023mxc}, we can see that the angular correlation pattern of SGWB caused by massive gravity is different from those caused by spin-2 ULDM.
If there is spin-2 ULDM signal in observational data of PTA, the detection of an anomalous shape of angular correlation curve at a specific frequency compared with curves at other frequencies would certainly help a lot with the detection of spin-2 ULDM effects. Based on these considerations, one strategy is to scan over the frequencies and characterize the deviation from the Hellings-Downs curve. 
For example, from the data set of CPTA \cite{Xu:2023wog}, the angular correlation curves are fitted with the data at a fixed frequency. Then we can choose certain frequencies at which the deviation maximizes, and fit such data with the deformed Hellings-Downs curve in our result \eqref{Gammaeff}.  Besides, the second data release from EPTA shows considerable ability to constrain the mass parameter space for scalar ULDM~\cite{EuropeanPulsarTimingArray:2023egv}. As in our case, it will help to constrain the parameter space for both mass and coupling constant $\alpha$ for spin-2 ULDM. Also, from Fig.~\ref{fig:eff10-6} to Fig.~\ref{fig:eff10-5}, we can see that with a larger coupling constant $\alpha$ of the spin-2 ULDM, the deformation is larger and easier to distinguish. In future work, we will constrain and fit the theoretical results with the observational data set.

\begin{acknowledgments}
This work is supported by the National Key Research and Development Program of China (Grants No. 2021YFA078304, No.2021YFC2201901) and the National Natural Science Foundation of China (Grants No. 12375059, No.12235019, No.11991052, No.11947302, and No.11821505).
\end{acknowledgments}

\appendix

\section{Theoretical model of  spin-2 ULDM}\label{app1}

The theory of massive spin-2 ULDM can originate from bimetric theory~\cite{Hassan:2011zd, Schmidt-May:2015vnx}, which was constructed originally in order to generalize Fierz-Pauli massive gravity in a non-linear way. Later it was found that this theory could provide a reasonable origin for spin-2 ULDM~\cite{Babichev:2016hir,Aoki:2016zgp, Marzola:2017lbt}. The total action is given by 
\begin{align}
  &  S  =\frac{{\mpl^2}}{1+\alpha^2}
    \int d^4x\Big[
    \sqrt{|\tg|}R(\tg)+\alpha^2\sqrt{|f|}R(f)\nonumber\\
    & -2\frac{\alpha^2{\mpl^2}}{1+\alpha^2}\sqrt{|\tg|}V(\tg,f;\beta_n)
   \Big]
    +\int d^4x\sqrt{|\tg|}{\cal L}_{\rm m}(\tg,\Psi), \label{action1}
\end{align}
where $M_{\rm Pl}$ is the reduced Planck mass, $f_{\mu\nu}$ is introduced as a new spin-2 field, while $\alpha$ is a dimensionless constant to regulate the difference of interactions for two spin-2 fields, and $V(\tg,f;\beta_n)$ stands for interaction terms between the two fields in order to avoid ghosts~~\cite{Hassan:2011zd,Schmidt-May:2015vnx,Babichev:2016hir}. This action can be linearized by considering perturbations
\begin{align}
    \tg_{\mu\nu}&=\bar{g}_{\mu\nu}+  \frac{1}{M_{\rm Pl}} (\G_{\mu\nu}-\alpha M_{\mu\nu})\ ,\\
    f_{\mu\nu}&=\bar{g}_{\mu\nu}+  \frac{1}{M_{\rm Pl}}(\G_{\mu\nu}+{\alpha^{-1}}M_{\mu\nu})\ ,
\end{align}
where $\bar{g}_{\mu\nu}$ is the background metric,  $\G_{\mu\nu}$ and $M_{\mu\nu}$ are the small perturbations. Furthermore, the resulting action can be diagonalized by this linear combination of the two metric perturbations.
Then the quadratic part of the total action \eqref{action1} becomes
\begin{align}
S^{(2)}=&\int d^4x \sqrt{|\bar{g}|}\Big[{\cal L}^{(2)}_{\rm GR}(\G)+{\cal L}^{(2)}_{\rm FP}(M)\nonumber\\
&\qquad -\frac{1}{M_{\rm Pl}}(\G_{\mu\nu}-\alpha M_{\mu\nu})T^{\mu\nu}(\Psi)\Big]\ ,
\end{align}
where ${\cal L}^{(2)}_{\rm GR}(X)$ is 2nd-order perturbation expansion of Einstein-Hilbert action, described by the Lichnerowicz operator, as
\begin{align}
    {\cal L}^{(2)}_{\rm GR}(X)=&-\frac{1}{2}{\mpl^2}X^{\mu\nu}{\cal E}^{~~\lambda\kappa}_{\mu\nu}X_{\lambda\kappa}\nonumber\\
    =&-\frac{1}{4}{\mpl^2}X^{\mu\nu}\Big(
    \delta^\lambda_\mu
\delta^\kappa_\nu\square-\bar{g}_{\mu\nu}\bar{g}^{\lambda\kappa}\square+\bar{g}^{\lambda\kappa}\nabla_\mu\nabla_\nu +\nonumber\\
    &+\bar{g}_{\mu\nu}\nabla^\lambda\nabla^\kappa-2\nabla^\lambda\nabla_{(\mu}\delta^\kappa_{\nu)}
    \Big)X_{\lambda\kappa}\ ,
\end{align}
while ${\cal L}^{(2)}_{\rm FP}(M)$ represents the Fierz-Pauli Lagrangian, which describes a massive spin-2 field
\begin{align}
    {\cal L}^{(2)}_{\rm FP}(M)={\cal L}^{(2)}_{\rm GR}(M)-\frac{m^2}{4}(M_{\mu\nu}M^{\mu\nu}-M^2)\label{lforEOM}\ ,
\end{align}
and $m=\sqrt{\beta_1+2\beta_2+\beta_3}M_{\rm Pl}$ can be identified as the mass for the mass eigenstate $M_{\mu\nu}$. Terms in $\G_{\mu\nu}$ can be combined with background metric  ${\calG}_{\mu\nu}=\bar{g}_{\mu\nu}+\frac{1}{M_{\rm Pl}}\G_{\mu\nu}$ and re-summed to recover Einstein-Hibert action. The resulting total action is
\begin{align}
    S_{\rm spin-2}=&{\mpl^2}\int d^4x \sqrt{|{\calG}|}
\left[R({\calG})+{\cal L}^{(2)}_{\rm FP}(M)+{\cal O}(M^3_{\mu\nu})\right] .
\end{align}

\end{document}